# Spin Seebeck effect in iron oxide thin films: Effects of phase transition, phase coexistence, and surface magnetism

Amit Chanda[1,*], Derick DeTellem[1], Yen Thi Hai Pham[1], Jenae E. Shoup[1], Anh Tuan Duong[2], Raja Das[2], Sunglae Cho[3], Dmitri V. Voronine[1], M. Tuan Trinh[1], Dario A. Arena[1], Sarath Witanachchi[1], Hariharan Srikanth[1,*], and Manh-Huong Phan[1,*]

[1] Department of Physics, University of South Florida, Tampa, Florida 33620, USA

[2] Faculty of Materials Science and Engineering, Phenikaa University, Yen Nghia, Ha-Dong District, Hanoi, 12116, Viet Nam

[3] Department of Physics and Energy Harvest-Storage Research Center, University of Ulsan, Ulsan 680-749, Republic of Korea

**Abstract**

Understanding impacts of phase transition, phase coexistence, and surface magnetism on the longitudinal spin Seebeck effect (LSSE) in a magnetic system is essential to manipulate the spin to charge current conversion efficiency for spincaloritronic applications. We aim to elucidate these effects by performing a comprehensive study of the temperature dependence of LSSE in biphase iron oxide (BPIO = $\alpha$-$Fe_2O_3$ + $Fe_3O_4$) thin films grown on Si (100) and $Al_2O_3$ (111) substrates. A combination of temperature-dependent anomalous Nernst effect (ANE) and electrical resistivity measurements show that the contribution of ANE from the BPIO layer is negligible compared to the intrinsic LSSE in the Si/BPIO/Pt heterostructure even at room temperature. Below the Verwey transition of the $Fe_3O_4$ phase, the total signal across BPIO/Pt is dominated by the LSSE. Noticeable changes in the intrinsic LSSE signal for both Si/BPIO/Pt and $Al_2O_3$/BPIO/Pt heterostructures around the Verwey transition of the $Fe_3O_4$ phase and the antiferromagnetic (AFM) Morin transition of the $\alpha$-Fe$_2$O$_3$ phase are observed. The LSSE signal for Si/BPIO/Pt is found to be almost two times greater than that for Al$_2$O$_3$/BPIO/Pt, an opposite trend is observed for the saturation magnetization though. Magnetic force microscopy reveals the higher density of surface magnetic moments of the Si/BPIO film compared to the Al$_2$O$_3$/BPIO film, which underscores a dominant role of interfacial magnetism on the LSSE signal and thereby explains the larger LSSE for Si/BPIO/Pt.

**KEYWORDS:** Iron oxide thin film; Magnetism; Spin Seebeck effect; Inverse spin Hall Effect; Spincaloritronics; Anomalous Nernst effect

*Corresponding authors: achanda@usf.edu (A.C.); sharihar@usf.edu (H.S.); phanm@usf.edu (M.H.P.)

# 1. INTRODUCTION

The past few years have witnessed a rapid growth in the field of generation and manipulation of pure spin currents for cutting edge spintronic applications.[1] The propagation of pure spin currents in a magnetic insulator (MI) in its magnetically saturated state takes place via collective excitation of the polarized spins which is also known as magnon.[2] Magnon spin current in a MI is beneficial over charge current in an ordinary ferromagnetic metal/semiconductor since (1) magnon spin current does not cause energy dissipation due to ohmic losses as electronic degrees of freedom become frozen, and (2) the spin diffusion length of magnon driven spin currents is much larger than that of charge currents.[3] From a green, efficient-energy harvesting perspective, thermally excited spin transport, also known as the spin Seebeck effect (SSE)[4], appears to be an effective way of generating incoherent magnon excitations in the THz regime[5]. The longitudinal spin Seebeck effect (LSSE) refers to the generation of a magnon spin current in a MI parallel to the direction of the applied temperature gradient.[5] These thermally generated magnons accumulate at the cold side of the thermal gradient[6, 7]. When a thin layer of heavy a metal (HM) with strong spin-orbit coupling (SOC), *e.g.*, platinum (Pt) is placed in close contact with the MI, the accumulated magnons at the interface between the MI and the HM layer drives pure spin current into the HM layer which is then converted into charge current via the inverse spin Hall effect (ISHE);[8] the generated electric voltage measured transverse to both the applied magnetic field direction and the temperature gradient is called the LSSE signal. The bilayer structure consisting of ferrimagnetic insulator $Y_3Fe_5O_{12}$ (YIG) and Pt is known as the hallmark system for LSSE.[9-13] Apart from insulating iron garnets,[14] insulating spinel ferrites *e.g.*, $CoFe_2O_4$,[15] have also emerged as another promising class of magnetic insulators for LSSE. Over the years, research has shown that the observation of LSSE is not only restricted to magnetic insulators, including ferrimagnetic and antiferromagnetic insulators[9, 16], but it has also been observed in ferromagnetic metals,[4] half-metallic ferromagnets, *e.g.*, mixed valent manganites[17] and ferromagnetic semiconductors, *e.g.*, GaMnAs[18]. However, for a bilayer made of ferromagnetic metal or, semiconductor and Pt, the measured LSSE voltage is usually contaminated by the spin polarized charge current driven by the temperature gradient across the ferromagnetic layer due to the anomalous Nernst effect (ANE),[19] as well as the proximity induced ANE in the Pt layer[20]. A quantitative disentanglement of these spurious signals is therefore necessary in order to extract the *intrinsic* LSSE contribution.[19] Other effects such as electronically/magnetically coupled structural phase transition,[21] phase coexistence,[22, 23] and surface magnetization[12, 13, 24-26] on the SSE have also been reported in various systems, but are yet to be understood. It is therefore essential to find a magnetic system within which the effects can be investigated systematically.

Magnetite ($Fe_3O_4$) is a unique half-metallic magnetic material[27] for LSSE investigations because of its famous Verwey transition ($\approx$ 120 K) at which the system undergoes a transformation from the high temperature conducting state to the low temperature insulating state. Ramos *et al.* have extensively investigated the LSSE in epitaxial $Fe_3O_4$/Pt bilayers,[28, 29] as well as in $[Fe_3O_4/Pt]_n$ multilayers,[30] and identified the ANE-free intrinsic LSSE contribution below the Verwey transition. They showed that both the magnetic proximity-induced ANE and the inherent ANE contribution of the magnetic layer were strongly suppressed in the Pt layer, and that the LSSE played a dominant role for the $Fe_3O_4$/Pt bilayer and $[Fe_3O_4/Pt]_n$ multilayers. However, there is a need for a quantitative disentanglement of the inherent LSSE, the ANE of the magnetic layer, and the proximity-induced ANE in the Pt layer to understand the temperature evolution of the *intrinsic* LSSE of $Fe_3O_4$/Pt in the vicinity of the Verwey transition. While the conductivity mismatch between the $Fe_3O_4$ and Pt layers has been shown to affect the intrinsic LSSE in $Fe_3O_4$/Pt, a clear understanding of the role of interfacial magnetism or surface magnetization of the $Fe_3O_4$ layer on the LSSE signal is lacking. Using scanning tunneling microscopy (STM), Condon *et a*l. observed surface reconstruction of a reduced $Fe_3O_4$ and revealed biphase ordering of the surface consisting of superlattices of $Fe_3O_4$ and $Fe_{1-x}O$ islands.[31, 32] Later, Lanier *et al.*[33] evidenced the coexistence of ferrimagnetic (FM) $Fe_3O_4$ and antiferromagnetic (AFM) $\alpha$-$Fe_2O_3$ islands instead of the $Fe_{1-x}O$ phase on the $\alpha$-$Fe_2O_3$ surface. Most interestingly, Genuzio *et al.*[34] observed transformation of the $Fe_3O_4$ phase into $\alpha$-$Fe_2O_3$ phase in $Fe_3O_4$ thin films grown on Pt and Ag substrates by annealing them not only in $O_2$ environment but also in ultrahigh vacuum (UHV). The $Fe_3O_4 \rightarrow \alpha$-$Fe_2O_3$ conversion in UHV which was observed only for the films grown on Pt substrate, is driven by the diffusion of Fe from the surface into the bulk of the film indicating the important role of the substrate on the formation of the bi-phase. These studies raise an interesting, fundamental question regarding the effect of phase coexistence (FM $Fe_3O_4$ and AFM $\alpha$-$Fe_2O_3$ phases) on the thermally driven spin transport in such iron oxide systems, which was not examined in the previous studies[28-30, 35].

In this context, we have chosen a biphase iron oxide (BPIO = $\alpha$-$Fe_2O_3$ + $Fe_3O_4$) film as a model system for probing effects of the phase transition and phase coexistence on the LSSE over a wide temperature range of 10 – 300 K. For studying the effect of surface magnetization, we measured LSSE in iron oxide thin films grown on two different substrates, Si (100) and $Al_2O_3$ (111), using molecular beam epitaxy. A combination of temperature-dependent ANE and electrical resistivity measurements on the BPIO layer shows the negligible contribution of the ANE from the BPIO layer as compared to the intrinsic LSSE signal in the Si/BPIO/Pt heterostructure even at room temperature. Below the Verwey transition, the total signal across BPIO/Pt is dominated by the LSSE. The Si/BPIO/Pt and Al$_2$O$_3$/BPIO/Pt heterostructures show noticeable changes in the LSSE signal around the Verwey transition of the Fe$_3$O$_4$ phase and the AFM Morin transition[36-38] of the $\alpha$-Fe$_2$O$_3$ phase, highlighting the important effects of the phase transition and coexistence. The LSSE signal for Si/BPIO/Pt is found to be almost two times higher than that for Al$_2$O$_3$/BPIO/Pt, due to the higher density of surface magnetic moments of the Si/BPIO film compared to the Al$_2$O$_3$/BPIO film. These findings underscore the dominating role of interfacial magnetism of the FM layer on the LSSE signal in FM/HM bilayers.

## 2. RESULTS AND DISCUSSION

Figure 1(a) shows the XRD patterns of the iron oxide films grown on Si (100) and Al$_2$O$_3$ (111) substrates. In addition to the cubic Fe$_3$O$_4$ phase[39, 40], the peaks associated with a secondary $\alpha$-Fe$_3$O$_4$ phase[41-43] are clearly visible in the XRD patterns for both Si/Fe$_3$O$_4$ and Al$_2$O$_3$/Fe$_3$O$_4$ films. The Raman spectra of both the iron oxide films also show characteristic peaks for both Fe$_3$O$_4$ and α-Fe$_2$O$_3$ phases (Figure 1(b)). For the Al$_2$O$_3$/Fe$_3$O$_4$ film, the α-Fe$_2$O$_3$ phase is characterized by six strong peaks at 229 cm$^{-1}$ ($A_{1g}$), 244 cm$^{-1}$ ($E_g$), 291 cm$^{-1}$ ($E_g$), 406 cm$^{-1}$ ($E_g$), 499 cm$^{-1}$ ($A_{1g}$), and 606 cm$^{-1}$ ($E_g$), and the Fe$_3$O$_4$ phase is characterized by two strong peaks at 535 cm$^{-1}$ and 665 cm$^{-1}$.[44-48] For the Si/Fe$_3$O$_4$ film, the 499 cm$^{-1}$ ($A_{1g}$) and 535 cm$^{-1}$ peaks associated with α-Fe$_2$O$_3$ and Fe$_3$O$_4$ phases, respectively are masked by the strong Si (520 cm$^{-1}$) peak. However, the remaining peaks associated with both the α-Fe$_2$O$_3$ and Fe$_3$O$_4$ phases are visible for the Si/Fe$_3$O$_4$ film. Some of the $E_g$ peaks associated with α-Fe$_2$O$_3$ are blue shifted possibly because of the strong Si-background. These results indicate that both Si/Fe$_3$O$_4$ and Al$_2$O$_3$/Fe$_3$O$_4$ films are not single phase but consist of two different phases of iron oxide. We designate the "Si/Fe$_3$O$_4$" and "Al$_2$O$_3$/Fe$_3$O$_4$" films as "Si/BPIO" and "Al$_2$O$_3$/BPIO" films, respectively throughout the manuscript, where BPIO represents a combination of Fe$_3$O$_4$ and $\alpha$-Fe$_3$O$_4$ phases. Figure 1(c) and (d) demonstrate the surface morphology of the Al$_2$O$_3$/BPIO and Si/BPIO films, respectively, characterized by the FESEM. It is evident that the crystallites are smaller in size (average size ~ 100 nm) and more closely packed for the Si/ BPIO film in comparison to the Al$_2$O$_3$/BPIO film. From the cross-sectional FESEM image of the Si/BPIO film (Figure 1(e)), the interfacial boundary between the BPIO film and Si substrate is clearly distinguishable. We performed the cross-sectional FESEM for all the four edges of the Si/BPIO bilayer structure and confirmed that the thickness of the BPIO layer is $\approx 80 \pm 5$ nm indicating the uniform growth of the BPIO film on the Si (100) substrate.

LSSE measurements on the Si/BPIO/Pt and Al$_2$O$_3$/BPIO/Pt heterostructures were performed. Figure 2(a) illustrates a schematic sketch of the LSSE measurement on the Si/BPIO/Pt heterostructure. An out-of-plane temperature gradient was applied along the +z-direction that generates a temperature difference, $\Delta T = +10$ K between the top (cold) and bottom (hot) surfaces of the heterostructure. An in-plane voltage was developed along the y-direction ($V_y$) in the Pt layer due to the ISHE, while scanning the external in-plane magnetic field applied along the x-direction. According to the bulk magnon theory, the concurrent presence of a temperature gradient ($\overrightarrow{\nabla T}$) and an in-plane dc magnetic field develops a spatial gradient of magnon accumulation in the magnetic layer (BPIO, for our case).[6, 7, 49-51] These accumulated magnons close to the BPIO/Pt interface are responsible for the transverse spin current pumping into the Pt layer.[7, 49, 51] The density of spin current ($\overrightarrow{J_S}$) at the Si/BPIO interface is related to the applied temperature gradient ($\overrightarrow{\nabla T}$) through the relation, $\overrightarrow{J_S} \propto -Q_{LSSE}\overrightarrow{\nabla T}$; where, $Q_{LSSE}$ is the spin Seebeck coefficient.[7, 52] This spin current enters the Pt layer and gets converted into charge current, $\overrightarrow{J_C} = \left(\frac{4\pi e}{h}\right)\theta_{SH}^{Pt}(\overrightarrow{J_S} \times \overrightarrow{\sigma})$; where, $\theta_{SH}^{Pt}$ is the spin Hall angle of Pt and $\vec{\sigma}$ is the spin-polarization vector.[53] The corresponding spin Seebeck voltage can be expressed as,[7, 54]

$$V_{LSSE} = R_y L_y \lambda_{Pt} \left(\frac{4\pi e}{h}\right) \theta_{SH}^{Pt} J_S \tanh\left(\frac{t_{Pt}}{2\lambda_{Pt}}\right), \quad (1)$$

where, $e, \hbar, R_y, L_y, \lambda_{Pt},$ and $t_{Pt}$ are the electronic charge, the reduced Planck's constant, the resistance between the contact probes on the Pt layer, the distance between the contact leads, the spin diffusion length, and the thickness of the Pt layer, respectively.

Since Fe$_3$O$_4$ is a half-metallic ferromagnet with a high Curie temperature ($T_C \sim 850$ K), it has high spin polarization of conduction electrons at room temperature. So, in presence of the out-of-plane temperature gradient and in-plane magnetic field, a transverse spin-polarized current is generated in the BPIO layer by the ANE; this gives rise to an extra voltage contribution ($V_{FM}^{ANE}$) to the total signal measured across the top Pt layer of the Si/BPIO/Pt heterostructure.[19] Under the application of a temperature gradient $\vec{\nabla T}$, the ANE voltage generated in a magnetic conductor with magnetization $\vec{M}$ is described by the cross product of $\vec{M}$ and $\vec{\nabla T}$ as,[55]

$$\vec{E_{ANE}} \propto Q_{ANE}(\mu_0 \vec{M} \times \vec{\nabla T}) \quad (2)$$

where, $\vec{E_{ANE}}$ is the electric field generated by ANE and $Q_{ANE}$ is the anomalous Nernst coefficient. Moreover, there is an additional voltage contribution ($V_{Prox}^{ANE}$) due to the magnetic proximity-induced ANE in the non-magnetic Pt layer.[19, 56] Thus for the configuration illustrated in Figure 2(a) with Pt on top, the LSSE voltage is entangled with the ANE contribution from the magnetic BPIO layer and the proximity induced ANE in the Pt layer. However, both the ANE voltages: $V_{FM}^{ANE}$ and $V_{Prox}^{ANE}$ are strongly suppressed due to the placement of the 5 nm thick top Pt layer.[28] Now, let us discuss about the contribution from the magnetic proximity-induced LSSE in the Pt layer. Because of the magnetic proximity, only few layers of Pt (with thickness, $d_{Pt}^{mag}$) close to the BPIO/Pt interface are magnetized, whereas the remaining layers (with thickness, $d_{Pt}^{NM}$) are unmagnetized. In presence of the temperature gradient, vertical spin currents are generated inside the magnetized Pt layer.[28] This spin current induces an additional in-plane charge current to the magnetic Pt/nonmagnetic Pt interface via the ISHE, contributing to the total signal with the voltage component, $V_{Prox,Pt}^{LSSE}$. Therefore, the total voltage measured across the Pt layer of the Si/BPIO/Pt heterostructure in the presence of the out-of-plane temperature gradient and in-plane applied magnetic field can be written as $V_y^{ANE+LSSE} = V_y^{LSSE} + V_{FM, Sup}^{ANE} + V_{Prox, Sup}^{ANE} + V_{Prox,Pt}^{LSSE}$, where $V_{FM, Sup}^{ANE}$ and $V_{Prox, Sup}^{ANE}$ are the suppressed ANE voltages due to the BPIO layer and the proximity-induced ANE voltage in the non-magnetic Pt layer. Bougiatioti *et al.* demonstrated that the magnetic proximity induced ANE signal in the non-magnetic Pt layer is significant for $Ni_{33}Fe_{67}$ (metallic)/Pt bilayer but negligible for $NiFe_2O_x$ (semiconducting)/Pt and becomes zero for $NiFe_2O_4$ (insulating)/Pt.[19] Since $Fe_3O_4$ is semiconducting at room temperature (with resistivity ~$10^{-2}$ Ω cm) and becomes insulating below the Verwey transition, the contribution of the proximity-induced ANE in the Pt layer can be neglected (previous report[28] shows that $\frac{V_{FM, Sup}^{ANE}}{\Delta T} \approx 7.5$ nV/K at $T = 300$ K considering fully magnetized one monolayer of Pt) in our case throughout the measured temperature range. Moreover, it is known that the LSSE voltage reduces with decreasing thickness of the magnetic layer.[10] Since $d_{Pt}^{mag}$ is very small, the magnetic proximity-induced LSSE contribution of the Pt layer, $V_{Prox,Pt}^{LSSE}$ is negligible. Therefore, the total voltage measured across the Pt layer of the Si/BPIO/Pt heterostructure becomes $V_y^{ANE+LSSE} = V_{LSSE} + V_{FM, Sup}^{ANE}$. We show the magnetic field dependence of the total voltage, $V_y^{ANE+LSSE}(H)$ across the Pt layer of the Si/BPIO/Pt heterostructure at selected temperatures between $T = 295$ and 75 K in Figure 2(b). The maximum field range was limited to $\mu_0 H = \pm 0.3$ T and the temperature difference between the hot and cold plates was fixed at $\Delta T = +10$ K for all the measurements. The $V_y^{ANE+LSSE}(H)$ curves clearly show well-defined hysteresis loops at all temperatures.

To decouple the intrinsic LSSE contribution from the ANE contaminations, ANE measurements were separately performed on the Si/BPIO bilayer structure. Figure 2(c) shows a schematic illustration of the ANE measurement on the Si/BPIO bilayer, using the same experimental conditions as the LSSE measurement on the Si/BPIO/Pt heterostructure. We show the magnetic field dependence of the ANE voltage, $V_y^{ANE}(H)$ at selected temperatures between $T = 295$ and 75 K in Figure 2(d). Like the $V_y^{ANE+LSSE}(H)$ curves, $V_y^{ANE}(H)$ curves also exhibit well-defined hysteresis loops. In Figure 2(e), we compared the $V_y^{ANE+LSSE}(H)$ and $V_y^{ANE}(H)$ isotherms at $T = 295$ for. It is evident that the signal is strongly enhanced by ~ 2 times upon the placement of the Pt layer on the BPIO layer. In Figure 2(f), we compared the $V_y^{ANE+LSSE}(H)$ and $V_y^{ANE}(H)$ isotherms at $T = 105$ K, which is below the Verwey transition. Surprisingly, the (ANE+LSSE) signal is ~10 times higher than the ANE signal, which underscores a crucial role of electrical resistivity of the BPIO film in controlling the intrinsic LSSE signal. As already mentioned, the ANE signal is suppressed by the placement of the Pt layer on top of the BPIO layer. Considering a parallel circuit arrangement of the BPIO and Pt layers, the degree of suppression of the ANE signal is usually expressed as the ratio of the electrical conductance of the BPIO layer ($G_{\alpha-Fe_2O_3+Fe_3O_4} = G_{\text{BPIO}}$) and the Pt layer ($G_{Pt}$). The suppressed ANE voltage across the Pt layer of the Si/BPIO /Pt heterostructure can thus be expressed as[19, 28]

$$V_{FM, Sup}^{ANE} = \left(\frac{f}{1+f}\right) V_{FM}^{ANE} \qquad (3)$$

where $f = \frac{G_{BPIO}}{G_{Pt}} = \frac{\rho_{Pt}}{\rho_{BPIO}} \cdot \frac{t_{BPIO}}{t_{Pt}}$. Here, $\rho_{BPIO}$ and $\rho_{Pt}$ are the electrical resistivity of the BPIO and Pt layers, and $t_{BPIO}$ and $t_{Pt}$ are the thickness of the BPIO and Pt layers, respectively. We measured the temperature dependence of resistivity of both 80 nm BPIO and 5 nm Pt layers separately (see the left inset of Figure 3(e)). At room temperature, $\rho_{BPIO}$ ($\approx 10^{-2}$ $\Omega$ cm) is three orders of magnitude higher than $\rho_{Pt}$ ($\approx 10^{-5}$ $\Omega$ cm). Hence, $\frac{G_{BPIO}}{G_{Pt}} \ll 1$, indicating $V_{FM,\,Sup}^{ANE} \ll V_{FM}^{ANE}$. Thus, the observed signal across the Pt layer of the Si/BPIO/Pt heterostructure is mostly dominated by the contribution from the intrinsic LSSE of the BPIO layer. Because of the semiconducting nature of the BPIO film $\left(\frac{\partial \rho_{BPIO}}{\partial T} < 0\right)$, $\rho_{BPIO}$ increases with lowering temperature, whereas $\rho_{Pt}$ $\left(\frac{\partial \rho_{Pt}}{\partial T} > 0\right)$ decreases. Hence, the correction factor $\left(\frac{f}{1+f}\right)$ also decreases with decreasing temperature (see the right inset of Figure 3(e)), indicating a stronger contribution of the intrinsic LSSE of the BPIO layer at low temperatures. Using the temperature dependence of the correction factor, we determined the intrinsic LSSE voltage, $V_y^{LSSE} = V_y^{ANE+LSSE} - V_{FM,\,Sup}^{ANE}$ of the Si/BPIO/Pt heterostructure at different temperatures. Figure 3(a) depicts the magnetic field dependence of the LSSE voltage, $V_y^{LSSE}(H)$ at selected temperatures between $T = 295$ and 75 K. A well-defined hysteresis loop is evident for $V_y^{LSSE}(H)$ isotherms at all temperatures.

Figure 3(c) shows the temperature dependence of the background-corrected voltages, $V_i = \frac{V_y^i(\mu_0 H = +0.3\,T) - V_y^i(\mu_0 H = -0.3\,T)}{2}$ for the Si/BPIO/Pt heterostructure; where the index $i$ refers to the ANE voltage, total voltage (ANE + LSSE), and intrinsic LSSE voltage. It is evident that $V_{ANE+LSSE}$ is almost 2 times higher than $V_{ANE}$ throughout the measured temperature range and both $V_{ANE}$ and $V_{ANE+LSSE}$ decrease slowly with decreasing temperature up to $T_A = 200$ K below which both of them undergo a sudden drop at $T = 185$ K. $V_{ANE+LSSE}$ increases slowly below $T = 185$ K and shows a broad maximum around $T_V = 120$ K below which it decreases rapidly. On the other hand, a careful observation reveals that $V_{ANE}$ slightly increases below $T = 175$ K and shows a weak maximum around $T = 145$ K but drops drastically with further lowering the temperature. As previously anticipated, the effective ANE contribution across the Pt layer is negligible and hence the intrinsic LSSE of the BPIO layer plays the dominating role. Therefore, a small difference in magnitude is observed between $V_{LSSE}$ and $V_{LSSE+ANE}$ throughout the measured temperature range, and the difference is insignificant below $T_V = 120$ K. Nevertheless, the temperature dependence of $V_{LSSE}$ follows the same trend as $V_{LSSE+ANE}$. The broad maximum in $V_{LSSE}$ can be explained from the temperature dependence of the magnitude of spin polarization, $\vec{\sigma}$. Previous study on the Fermi edge spin polarization in $Fe_3O_4$ has shown that the magnitude of $\vec{\sigma}$ increases with decreasing temperature due to gradual opening of the energy gap between the minority $t_{2g}$ spin states and the majority $e_g$ spin states at the Fermi level and attains its maximum value near $T_V$ but decreases with further lowering temperature.[57] Since ISHE-induced charge current in the Pt layer is proportional to the cross product $(\vec{J_S} \times \vec{\sigma})$,[53] the LSSE voltage also shows a broad maximum around $T_V$ and decreases below $T_V$.

To correlate the temperature profile of thermo-spin transport in Si/BPIO/Pt heterostructure with magnetic properties of the BPIO layer, we investigated the temperature dependence of magnetization $M(T)$ of Si/BPIO in zero-field cooled (ZFC) and field-cooled (FC) mode (see Figure 3(d)) measured under an in-plane magnetic field of $\mu_0 H = 0.1$ T. Both ZFC and FC $M(T)$ curves exhibit a pronounced downturn around $T_V \sim 120$ K (determined from the maximum in the d$M$/d$T$ curve), which is identified as the Verwey transition of the $Fe_3O_4$ phase. In addition, with decreasing temperature from 300 K, there is a slow decrease in the magnetization below $T \approx 250$ K in the ZFC $M(T)$, which is followed by a faster decrease below $T_A \approx 200$ K and a pronounced dip at $T_D \sim 185$ K. This feature is identified as the Morin transition from the high temperature weakly FM to the low temperature AFM ordering of the $\alpha$-Fe$_2$O$_3$ phase.[36-38, 58, 59] Typically, the Morin transition is sharp with the critical temperature of $T_N^{\alpha-Fe_2O_3} \approx 250\ K$ and the magnetization-dip is usually observed above 200 K. However, in our case, the transition is rather broad and the magnetization-dip occurs around $T_D = 185$ K. It is reported that the Morin transition strongly depends on the $\alpha$-Fe$_2$O$_3$ film thickness and it shifts to lower temperatures with decreasing thickness.[59] We believe that our system is phase separated into multiple micro- or, nanoscale regions of $\alpha$-Fe$_2$O$_3$ and Fe$_3$O$_4$ with different sizes and hence different $T_N^{\alpha-Fe_2O_3}$ associated with the $\alpha$-Fe$_2$O$_3$ phases. Such a distribution of $T_N^{\alpha-Fe_2O_3}$ can broaden the average macroscopic Morin transition of the $\alpha$-Fe$_2$O$_3$ phase. Hence, the significant decrease in the temperature profiles of both $V_{ANE}$ and $V_{ANE+LSSE}$ below $T_A = 200$ K and the pronounced dip-like feature at $T_D = 185$ K are associated with the decrease in magnetization originated from the AFM Morin transition of the $\alpha$-Fe$_2$O$_3$ phase.

In order to understand whether the $\alpha$-Fe$_2$O$_3$ and Fe$_3$O$_4$ phases are exchange coupled in our films, we performed exchange bias (EB) measurements. The magnetic field dependent magnetization, $M(H)$, measurements were performed on the Si/BPIO film at $T = 10$ K after the sample was cooled from 300 K in the presence (FC) or in the absence (ZFC) of an in-plane magnetic field of $\mu_0H = \pm1$ T (see Figure 3(b)). We have observed a small but prominent shift in the FC $M(H)$ loops towards the negative (positive) $H$-axis for the cooling field $\mu_0H = +1$ T (-1 T) (for clarity, see an inset of Figure 3(b)), indicating the presence of an EB effect. Considering that the $M(H)$ loop intersects the positive and negative $H$-axis at $H^+$ and $H^-$, respectively, the EB field can be defined as, $H_{EB} = \left|\frac{H^+ + H^-}{2}\right|$. For the FC $M(H)$ curves at 10 K, $H_{EB} = 55$ Oe (62 Oe) for the cooling field $\mu_0 H = +1$ T (-1 T). Such a small EB field indicates weak interfacial exchange coupling between the $\alpha$-$Fe_2O_3$ and $Fe_3O_4$ phases and educates the existence of a region of disordered spins separating these two phases.[60, 61] In addition to the shift along the $H$-axis, the FC $M(H)$ loops also exhibit small vertical shifts with respect to the ZFC $M(H)$ loop, which is possibly related to the weak pinning of the disordered spins at the AFM $\alpha$-$Fe_2O_3$/FM $Fe_3O_4$ interface[60, 62-64] which became frozen at low temperatures when the sample was cooled down in zero magnetic field. Since the EB and pinning effects decrease significantly with increasing temperature,[65] these effects could be negligibly small in our temperature region of interest (75 K $\leq T \leq$ 300 K) for the thermal spin transport measurements. The main panel of Figure 3(e) exhibits the temperature dependence of electrical resistivity ($\rho$) of Si/BPIO, which clearly shows a significant increase in the $\rho$ value below the Verwey transition ($T_V \sim$ 120 K) indicating a gradual transformation from the metallic to insulating phase ($\Delta\rho \approx 5$ m$\Omega$. cm for 120 K $\leq T \leq$ 300 K, whereas $\Delta\rho \approx 15$ m$\Omega$ cm for 50 K $\leq T \leq$ 120 K). However, it is evident that $\rho(T)$ does not show a rapid increase below $T_V$, which is a signature feature for the metal-insulator transition of $Fe_3O_4$ but increases gradually. Earlier studies indicate that $\rho(T)$ for $\alpha$-$Fe_2O_3$ does not show any steep increase around $T_V \sim$ 120 K upon lowering temperature but increases gradually with decreasing temperature.[66] Since the electrical resistivity of the $\alpha$-$Fe_2O_3$ phase is higher than that of the $Fe_3O_4$ phase, the metal-to-insulator transition at $T_V$ associated with the $Fe_3O_4$ phase is masked by the contribution of the $\alpha$-$Fe_2O_3$ phase. Hence, the $\rho(T)$ behavior for our Si/iron oxide film agrees with the coexistence of $Fe_3O_4$ and $\alpha$-$Fe_2O_3$ phases. Thus, the dominating contribution of the intrinsic LSSE for our Si/BPIO/Pt heterostructure below the Verwey transition is consistent with both the magnetization and electrical transport results.

Most importantly, the LSSE voltage has contributions from both the $Fe_3O_4$ and $α$-$Fe_2O_3$ phases, which highlights the effect of phase coexistence on the LSSE signal.

Figure 4(a)-(c) compare the magnetic field dependence of $V_y^{ANE+LSSE}(H)$, $V_y^{ANE}(H)$ and $V_y^{LSSE}(H)$, respectively on the left-y scale (symbol) and corresponding magnetization isotherms, $M(H)$ on the right y-scale (solid line) at $T = 295$ K for the Si/BPIO/Pt heterostructure. It is evident that the hysteresis loops for $V_y^{ANE+LSSE}(H)$, $V_y^{ANE}(H)$ and $V_y^{LSSE}(H)$ closely match with that of the $M(H)$ curves reflecting the intrinsic origin of both LSSE and ANE signals above the Verwey transition. In other words, the thermo-spin transport observed in our Si/BPIO/Pt heterostructure is controlled by the magnetic properties of the BPIO film. In order to confirm the intrinsic origin of the thermo-spin transport properties below the Verwey transition, we also demonstrate the $V_y^{ANE+LSSE}(H)$, $V_y^{ANE}(H)$ and $V_y^{LSSE}(H)$ isotherms on the left-y scale (symbol) and corresponding $M(H)$ curves on the right y-scale (solid line) at $T = 75$ K in Figure 4(d)-(f), respectively. Similar to $T = 295$ K, the $V_y^{ANE+LSSE}(H)$, $V_y^{ANE}(H)$ and $V_y^{LSSE}(H)$ curves clearly mimic the hysteresis loop of the corresponding $M(H)$ curve, which unambiguously dictates that the thermo-spin transport properties of our Si/BPIO/Pt heterostructure are governed by the intrinsic magnetic properties of the BPIO film below the Verwey transition as well.

We have also performed LSSE measurements for different thermal gradient intervals, $\Delta T$. Figure 5(a) shows the magnetic field dependence of the total voltage, $V_y^{ANE+LSSE}(H)$ across the Pt layer of the Si/BPIO/Pt heterostructure for different values of $\Delta T$ between +5 and +15 K while keeping the sample temperature fixed at $T = 295$ K. A significant increase in the $V_y^{ANE+LSSE}$ value with increasing temperature gradient is noted. To have a better understanding, we have shown the two-dimensional surface plots of $V_y^{ANE+LSSE}(H)$ isotherms for unipolar field scan: $+\mu_0 H_{sat} \rightarrow -\mu_0 H_{sat}$ ($\mu_0 H_{sat}$ is the magnetic field required magnetically saturate the BPIO film) in Figure 5(b), which clearly demonstrates an enhancement in the (ANE+LSSE) signal with increasing $\Delta T$. In Figure 5(c), we plot the background-corrected value of $V_y^{ANE+LSSE}(\mu_0 H = 0.3\ T)$, defined as $V_{ANE+LSSE} = \frac{V_y^{ANE+LSSE}(\mu_0 H = +\ 0.3\ T) - V_y^{ANE+LSSE}(\mu_0 H = -\ 0.3\ T)}{2}$ as a function of $\Delta T$, which clearly demonstrates a linear behavior. Since we have already established that the contribution of ANE across the Pt layer of the Si/BPIO/Pt heterostructure is negligible even at room temperature, such a linear $\Delta T$-dependence of $V_{ANE+LSSE}$ reflects the linear $\Delta T$-dependence of the intrinsic LSSE signal originated from the thermally driven magnons in the BPIO layer, which is expected.

To further understand the role of the mixed phases ($Fe_3O_4 + \alpha\text{-}Fe_2O_3$ phase) on the LSSE signal, we performed thermo-spin transport measurements on iron oxide films grown on $Al_2O_3$ substrates, namely, the $Al_2O_3$/BPIO/Pt heterostructure. Figure 6(a) shows a schematic sketch of the LSSE measurement on the $Al_2O_3$/BPIO/Pt heterostructure. For this configuration, as already discussed, the intrinsic LSSE voltage is entangled with the ANE contribution from the BPIO layer. We demonstrate the magnetic field dependence of the total voltage, $V_y^{ANE+LSSE}(H)$ measured across the Pt layer of the $Al_2O_3$/BPIO/Pt heterostructure at selected temperatures between $T = 295$ and 75 K in Figure 6(b). Since we have already established that the ANE signal is largely suppressed in the Pt layer, we can say that the measured voltage across the Pt layer in the $Al_2O_3$/BPIO/Pt heterostructure is dominated by the intrinsic LSSE contribution. In Figure 6(c) and (d), we compare the two-dimensional surface plots of $V_y^{ANE+LSSE}(H)$ isotherms for $Al_2O_3$/BPIO/Pt and Si/BPIO/Pt heterostructures for unipolar field scan: $+\mu_0 H_{sat} \rightarrow -\mu_0 H_{sat}$. It is clear that, (i) similar to the Si/BPIO/Pt heterostructure, the total voltage signal undergoes major changes around $T_A \sim 200$ K, $T_D = 185$ K and $T_V = 120$ K for the Al$_2$O$_3$/BPIO/Pt heterostructure, (ii) the $V_y^{ANE+LSSE}(H)$ hysteresis loops for the Al$_2$O$_3$/BPIO/Pt heterostructure is narrower than that for the Si/BPIO/Pt heterostructure, and (iii) the magnitude $V_y^{ANE+LSSE}(H)$ signal for the Si/BPIO/Pt heterostructure is twice of that for the Al$_2$O$_3$/BPIO/Pt heterostructure.

Figure 6(e) demonstrates the magnetic field dependence of $V_y^{ANE+LSSE}(H)$ on the left-y scale (symbol) and corresponding magnetization isotherms, $M(H)$ on the right y-scale (solid line) at $T = 295$ K for the Al$_2$O$_3$/BPIO/Pt heterostructure. Like the Si/BPIO/Pt heterostructure, the hysteresis loop of $V_y^{ANE+LSSE}(H)$ also matches closely with that of the $M(H)$ curve indicating a strong correlation between the thermo-spin transport in our Al$_2$O$_3$/BPIO/Pt heterostructure and the magnetic properties of the BPIO film. In Figure 6(f) compares the temperature dependence of the background-corrected voltage normalized with respect to the distance between the contact leads ($L_y$) on the Pt layer, $\frac{V_{ANE+LSSE}}{L_y} = \frac{V_y^{ANE+LSSE}(\mu_0 H = +0.3\,T) - V_y^{ANE+LSSE}(\mu_0 H = -0.3\,T)}{2L_y}$ at $\mu_0 H = 0.3\,T$ for Al$_2$O$_3$/BPIO/Pt and Si/BPIO/Pt heterostructures. It is evident that $\frac{V_{ANE+LSSE}}{L_y}(T)$ for both the heterostructures follows nearly the same trend; upon decreasing the temperature from $T = 300$ K, $V_{ANE+LSSE}$ for both the heterostructures decreases considerably below $T_A \sim 200$ K, then shows a pronounced dip at $T_D = 185$ K, and then a broad maximum around $T_V = 120$ K followed by a drastic decrease in the voltage signal at low temperatures. However, $\frac{V_{ANE+LSSE}}{L_y}$ for Si/BPIO/Pt is nearly 2 times higher than that for Al$_2$O$_3$/BPIO/Pt throughout the measured temperature range. To correlate the temperature profile of thermo-spin transport in Al$_2$O$_3$/BPIO/Pt heterostructure with magnetic properties of the BPIO layer, we have also investigated ZFC and FC $M(T)$ for the Al$_2$O$_3$/BPIO measured under the application of an in-plane magnetic field μ$_0$H = 0.1 T (see Figure 6(g)). Similar to the Si/BPIO film, both ZFC and FC $M(T)$ curves for Al$_2$O$_3$/BPIO exhibit a pronounced downturn around the Verwey transition, $T_V$ ~ 120 K. Moreover, the ZFC $M(T)$ also exhibits a pronounced dip around $T_D$ = 185 K, indicating the coexistence of the FM α-Fe$_3$O$_4$ and AFM α-Fe$_2$O$_3$ phases. Note that the saturation magnetization of the Al$_2$O$_3$/BPIO film is almost two-times higher than that of the Si/BPIO film, whereas the LSSE signal for Si/BPIO/Pt is nearly twice of that for Al$_2$O$_3$/BPIO/Pt. Such a discrepancy highlights that the thermal spin injection across the BPIO/Pt junction is dominated by the interfacial magnetic properties (or surface magnetization of the BPIO layer) rather than the bulk magnetic properties.[13, 24, 25, 67, 68] Although the AFM α-Fe$_2$O$_3$ phase is present in both Si/BPIO and Al$_2$O$_3$/BPIO films, the higher value of LSSE voltage observed in Si/BPIO/Pt as compared to Al$_2$O$_3$/BPIO/Pt is possibly due to the higher value of surface magnetization, *i.e.*, the higher density of surface spins of the BPIO layer for the Si/BPIO/Pt heterostructure, which highlights the important role of the substrate in manipulating the surface spin density and hence the thermal spin injection across the BPIO/Pt interface.

To elucidate this, we have studied the surface magnetic domain structures of the Fe$_3$O$_4$ films grown on Si and Al$_2$O$_3$ substrates, using the AFM/MFM technique. Figure 7(a) and (c) show the AFM images of the Si/BPIO and Al$_2$O$_3$/BPIO films, respectively. The height contrast profiles clearly reflect that the BPIO films grown on Si (100) substrate have less surface roughness than those grown on Al$_2$O$_3$ (111) substrate. The MFM images reveal the surface magnetic domain structures of the Si/BPIO (Figure 7(b)) and Al$_2$O$_3$/BPIO films (Figure 7(d)), respectively. Irregular domains are clearly noticeable for the Si/BPIO surface, whereas those are obscure for the Al$_2$O$_3$/BPIO film. The superimposition of MFM (Figure 7(c)) over AFM (Figure 7(d)) images of the Al₂O₃/BPIO film reveals that the phase shift obtained in the lift mode scan closely follows the footprint of the topography scan. In other words, the uneven topographical surface, rather than the surface magnetic moments, mainly contributes to the phase difference in the magnetic probe oscillations. The significant discrepancy between the magnetic domain structures of the Si/BPIO and Al₂O₃/BPIO films suggests the higher density of surface magnetic moments (or the higher surface magnetization) on the Si/BPIO film as compared to the Al₂O₃/BPIO film. It is the higher interfacial spins density that amplifies spin current injection across the BPIO/Pt interface giving rise to the enhanced LSSE in the Si/BPIO/Pt heterostructure, as compared to the Al₂O₃/BPIO/Pt heterostructure. [25, 67, 68]

Now, let us understand why the surface magnetization or the surface spin density is a deciding factor for the enhanced spin current injection but not the bulk saturation magnetization. As shown before, the ISHE-induced charge current in the Pt layer depends on the cross-product $(\vec{J_S} \times \vec{\sigma})$. The spin polarization vector $\vec{\sigma}$ is proportional to the projection of the in-plane surface magnetization, $\vec{M_{Surf}^{IP}}$.[67, 68] Hence, the ISHE-induced charge current in the Pt layer can be expressed as $\vec{J_C} \propto \left|\vec{J_S} \times \vec{M_{Surf}^{IP}}\right|$, which clearly explains the higher LSSE signal observed in the Si/BPIO/Pt heterostructure in comparison to the Al₂O₃/BPIO/Pt heterostructure. To explicate further, let us recall the magnetization dynamics of the magnetic layer (BPIO) in presence of a temperature gradient, $\vec{\nabla T}$ and an applied magnetic field, $\vec{H_{DC}}$. For such configuration, the stochastic Landau Lifshitz Gilbert (LLG) equation in the framework of the linear response theory of the LSSE can be expressed as[69]

$$\frac{\partial \vec{M}}{\partial t} = -\gamma \vec{M} \times (\vec{H_{eff}} + \vec{h}) + \frac{\alpha}{M_S}\left(\vec{M} \times \frac{\partial \vec{M}}{\partial t}\right), \qquad (4)$$

where $\overrightarrow{H_{eff}}$ is the effective magnetic field consisting of $\overrightarrow{H_{DC}}$ and anisotropy field, $\vec{h}$ is the noise field which accounts for the thermal fluctuations; $\gamma$ and $\alpha$ are the gyromagnetic ratio and the Gilbert's damping constant, respectively. As discussed before, the magnon spin current pumping from the magnetic layer (BPIO) into the Pt layer depends on the magnon accumulation close to the BPIO/Pt interface.[7] However, the magnon accumulation in the magnetic film is controlled by a critical propagation length ($\xi$) for the thermally generated magnons.[10, 70] Only those magnons contribute to the LSSE signal that are excited at a distance $\xi$ from the BPIO/Pt interface, where, $\xi$ is the average magnon propagation length of the thermally excited magnons.[10, 70] For magnetic materials with low saturation magnetization ($M_S$), the second term (damping term) in Eqn. (4) becomes significant. Since $\xi$ is inversely proportional to $\alpha$, for a magnetic material with high spin wave damping, only the magnons which are close to the interface actively participate in spin current injection to the Pt layer.[25] A recent study[35] indicates that $\xi \approx 19$ nm at 300 K for $Fe_3O_4$ thin films which reduces to $\approx 13$ nm at 50 K, both of which is sufficiently smaller than that of YIG thin films ($\xi \sim$ 90-140 nm)[10]. Since our BPIO films possess multiple phases of iron oxides, the extrinsic contributions to $\alpha$ is expected to be more pronounced, which can lead to further lowering of the $\xi$ - value. Such low value of $\xi$ in our iron oxide thin films also implies that the surface magnetization plays leading role in controlling the LSSE signal in our BPIO films. Since (a) $M_S$ for the Si/BPIO film is lower than the $Al_2O_3$/BPIO film, and (b) the surface spin density is higher for the Si/BPIO film than the $Al_2O_3$/BPIO film, the spin current injection efficiency is higher for the Si/BPIO/Pt heterostructure compared to the $Al_2O_3$/BPIO/Pt heterostructure, which fully supports our observation. Our study indicates that the thermally generated spin current injection across a FM/HM interface and consequently the LSSE signal strength can be significantly enhanced by manipulating surface spin density via appropriate substrate selection. We have demonstrated that the Si (100) substrate is a good choice for interfacial magnetism engineering across the FM/HM interface, which will pave the way for the development of cost-effective and energy-efficient spincaloritronic devices.

## 3. CONCLUSIONS

We have systematically investigated the temperature dependence of LSSE in biphase iron oxide thin films grown on Si (100) and $Al_2O_3$ (111) substrates. We have demonstrated that the contribution of ANE from the BPIO layer is negligible compared to the intrinsic LSSE signal in the Si/BPIO/Pt heterostructure. Below the Verwey transition, the total signal across BPIO/Pt is dominated by the LSSE. We have observed the noticeable changes in the LSSE signal for both Si/BPIO/Pt and $Al_2O_3$/BPIO/Pt heterostructures around the Verwey transition of the $Fe_3O_4$ phase and the AFM Morin transition of the $\alpha$-$Fe_2O_3$ phase. Owing to the higher density of surface magnetic moments of the Si/BPIO film compared to the $Al_2O_3$/BPIO film, the LSSE signal for the Si/BPIO/Pt heterostructure is almost two times higher than that for $Al_2O_3$/BPIO/Pt, despite that the saturation magnetization is ~2 times higher in the latter than in the former. These findings underscore the significant effects of phase transition, phase coexistence and surface magnetism on the LSSE signal in FM/HM bilayers.

## 4. EXPERIMENTAL SECTION

Biphase iron oxide (BPIO) films of thickness ~ 80 nm were grown on Si (100) and $Al_2O_3$ (111) substrates of dimensions 5 x 5 x 0.5 $mm^3$ using molecular beam epitaxy (MBE) (VG Semicon, Inc.) technique. The substrates were initially cleaned with methanol before being loaded into the vacuum chamber. After pre-heating at 600℃ or 30 min in ultra-high vacuum (UHV), the substrates were cooled down to 400℃, and maintained at that temperature during the growth. Fe was evaporated from an effusion cell with the evaporation rate of 0.2 Å/s under oxygen pressure of $8.2 \times 10^{-6}$ Torr. A thin Platinum (Pt) layer of thickness ~ 5 nm was deposited on the BPIO films using DC sputtering. The crystal phase characterization of the BPIO thin films was performed using Bruker AXS powder X–ray diffractometer (XRD) with Cu–Kα radiation at room temperature. A Horiba LabRAM HR Evolution Raman system equipped with a 532 nm diode-laser with 0.15 mW power was used for the Raman spectrum acquisition. Field emission scanning electron microscopy (FESEM) was also carried out to observe microstructural variation in the films. The film thickness was measured by cross-sectional FESEM measurements. The in-plane (IP) static magnetic characterization of the samples was performed using a vibrating sample magnetometer (VSM) attached to the physical property measurement system (PPMS) (Quantum Design, Inc., USA). Electrical resistivity measurements of the films were performed using the DC transport option of the PPMS. Both atomic force microscopy (AFM) and magnetic force microscopy (MFM) images were obtained using a scanning probe microscopy (SPM) system (OmegaScope-R, AIST-NT), in tapping mode for sample surface topography and lift mode for magnetic domain observation. A MikcroMasch Co-Cr coated Si magnetic probe (225 x 27.5 x 3 mm$^3$, with tip radius < 60 nm) was used for the measurements. The force constant and the resonant frequency of the tip were ≈ 2.8 Nm$^{-1}$ and 75 kHz, respectively. The AFM/MFM images are of ~300 px x 300 px and scanning rate was 0.5 Hz.

The LSSE and ANE on the films were measured within the temperature range 75 K ≤ $T$ ≤ 295 K and magnetic fields up to $\mu_0 H$ = 0.3 T using a custom designed set up by making use of a universal sample puck for the PPMS. Both these measurements were performed using the longitudinal configuration. The heterostructures were sandwiched between two copper plates (hot and cold). The bare surface of each of the copper plates was covered with a thin layer of Kapton tape to electrically isolate the sample from the copper plates. To maintain a constant temperature difference between these two plates, temperatures of both these plates were controlled by the PID control technique using two temperature controllers. The cold (top) plate was thermally anchored to the PPMS puck base (heat bath) via Molybdenum screws and the hot (bottom) plate was thermally separated from the PPMS puck base by a thick Teflon block to maintain a temperature difference of ~ 10 K between the PPMS puck base and the hot plate. A temperature gradient was applied along the +z-direction using two Pt 100 RTD sensors (used as resistive heaters) attached to both these plates that generates a temperature difference of Δ$T$ = 10 ±0.005 K between these two plates. A calibrated Si-diode sensor was attached to each of these plates which were used as thermometers that precisely recorded the temperatures $T_{hot}$ and $T_{cold}$ corresponding to temperatures of the hot and cold plates, respectively. The sample temperature was estimated as the average temperature, $T_{sample} = \frac{T_{hot}+T_{cold}}{2}$. Cryogenic N-grease was applied to both the surfaces of the hot and cold plates to maintain good thermal connectivity with the sample surface. To maintain the same thermal contact conditions, the same amount of N-grease between the sample and the hot/cold plates was applied for all the samples. The ISHE voltage ($V_y$) across the Pt layer of the Si/BPIO/Pt and Al$_2$O$_3$/BPIO/Pt heterostructures was measured along the *y* direction using a Keithley 2182A nanovoltmeter, while sweeping a DC magnetic field applied along the *x* direction by the superconducting magnet of the PPMS. After setting the PPMS base temperature and the temperatures of the hot and cold plates to the desired values, temperature difference between these two plates were monitored using the LabVIEW and the magnetic field dependent LSSE/ANE measurements were started only when a stable temperature difference was attained.

**Author Contributions**

M.H.P., A.C., A.T.D., and R.D. initiated the concept. A.C. designed and conducted magnetic and spin-thermo-transport experiments, analyzed the data, and wrote the first draft. D.D. and A.C. performed and analyzed SEM, XRD, and Raman data. Y.T.H.P., M.T.T., and D.V.V. performed and analyzed MFM data. J.E.S. and D.A.A. deposited Pt on the BPIO films. A.T.D., R.D., and S.L.C. grew the BPIO films. All authors participated in discussions and finalizing the manuscript. M.H.P. led the project.

**ACKNOWLEDGEMENTS**

Research at the University of South Florida was supported by the U.S. Department of Energy, Office of Basic Energy Sciences, Division of Materials Sciences and Engineering under Award No. DE-FG02-07ER46438.

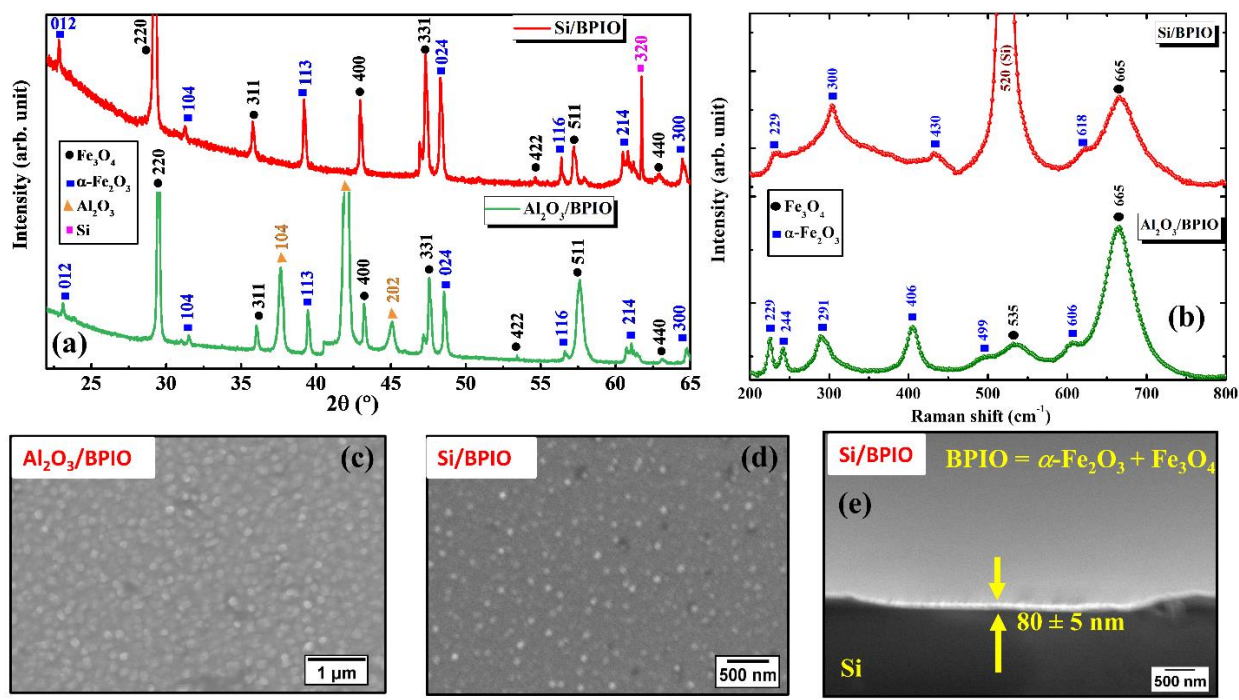

**Figure 1.** (a) Room temperature X-ray diffraction (XRD) pattern of the biphase iron oxide (BPIO) film grown on Si (100) and $Al_2O_3$ (111) substrates; peaks associated with the $Fe_3O_4$ and $\alpha\text{-}Fe_2O_3$ phases are evident, the orange circles represent peaks associated with the $Al_2O_3$ (111) substrate, (b) Raman spectra of the biphase iron oxide (BPIO) films grown on Si (100) and $Al_2O_3$ (111) substrates consisting of strong peaks associated with the $Fe_3O_4$ and $\alpha\text{-}Fe_2O_3$ phases, (c) and (d) FESEM images of the film surfaces of Si/BPIO and $Al_2O_3$/BPIO films, respectively. (e) Cross-sectional SEM image of the Si/BPIO film confirming that the average film thickness is ≈ 80 nm.

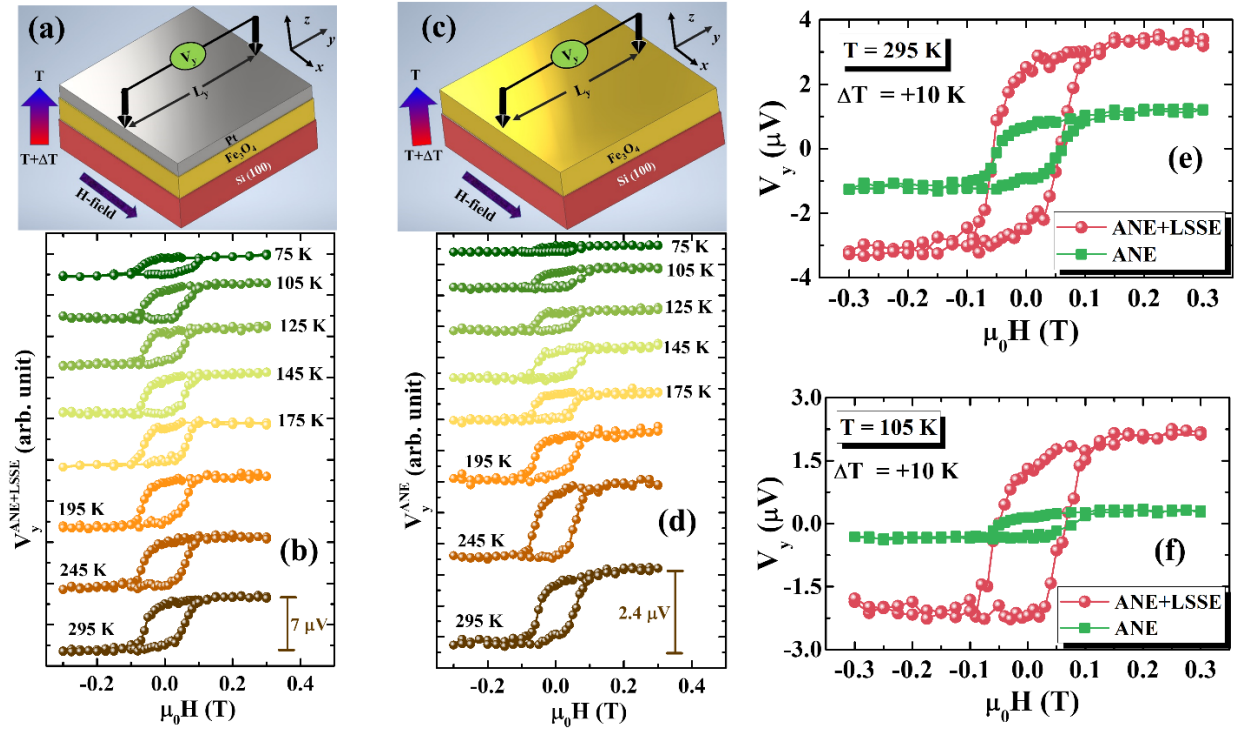

**Figure 2.** (a) and (c) Schematic illustrations of the longitudinal spin Seebeck effect (LSSE) and anomalous Nernst effect (ANE) measurements on the Si/BPIO/Pt and Si/BPIO heterostructures; (b) and (d) magnetic field dependence of the total voltage, $V_y^{ANE+LSSE}(H)$ measured across the Pt layer of the Si/BPIO/Pt heterostructure and ANE voltage, $V_y^{ANE}(H)$ measured across the bilayer Si/BPIO, respectively, at selected temperatures between $T = 295$ and 75 K for applied temperature gradient $\Delta T = +10$ K. Comparison of $V_y^{ANE+LSSE}(H)$ and $V_y^{ANE}(H)$ are shown at (e) $T = 295$ K (above the Verwey transition) and (f) T = 105 K (below the Verwey transition) for $\Delta T = +10$ K.

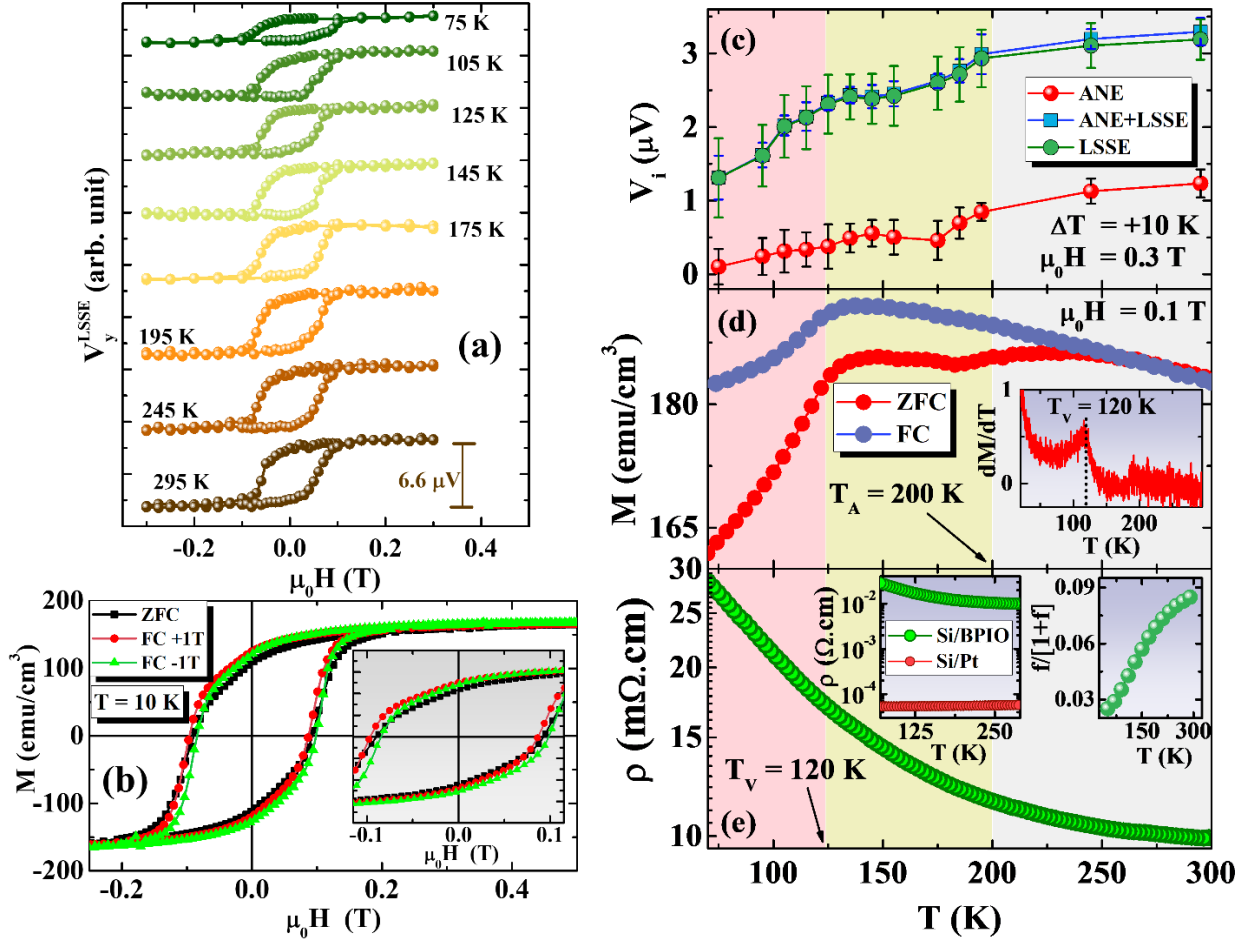

**Figure 3.** (a) Magnetic field dependence of the ANE-free LSSE voltage, $V_y^{LSSE}(H)$ of Si/BPIO/Pt heterostructure at selected temperatures between $T = 295$ and $75$ K; (b) main panel: magnetic field dependence of magnetization, M(H) of Si/BPIO bilayer measured after cooling the film in zero field (ZFC) as well as in $\mu_0H = +1$ T and $\mu_0H = -1$ T magnetic fields and inset shows the expanded view indicating small exchange bias effect. (c) Temperature dependence of the background-corrected voltages at $\mu_0H = 0.3$ T, $V_i = \frac{V_y^i(\mu_0H=+0.3\,T) - V_y^i(\mu_0H=-0.3\,T)}{2}$; where the index $i$ = ANE voltage, total voltage (ANE + LSSE), and intrinsic LSSE voltage. (d) Temperature dependence of the ZFC and FC magnetizations of Si/BPIO measured in a magnetic field of $\mu_0H = 0.1$; inset shows temperature dependence of the d$M$/d$T$ curve showing prominent maximum at the Verwey transition, $T_V \sim 120$ K. (e) Temperature dependence of electrical resistivity, $\rho(T)$ of Si/BPIO bilayer; left inset shows comparison of $\rho(T)$ for Si/BPIO and Si/Pt(5nm) bilayers and the right inset shows the temperature dependence of the ANE correction factor $\left(\frac{f}{1+f}\right)$.

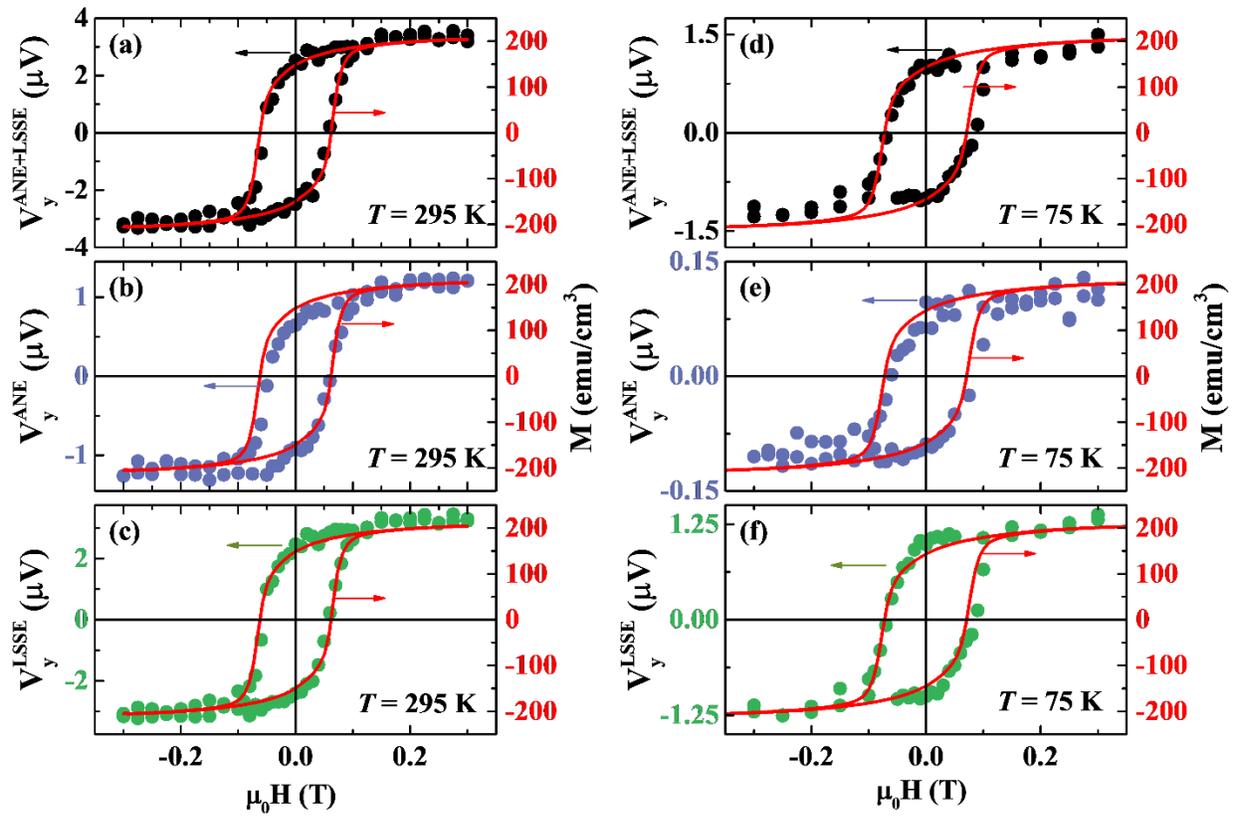

**Figure 4.** Comparison of the magnetic field dependence of $V_y^{ANE+LSSE}(H)$, $V_y^{ANE}(H)$ and $V_y^{LSSE}(H)$, respectively on the left-y scale (symbol) and corresponding magnetization isotherms, $M(H)$ on the right y-scale (solid line) at (a)-(c) $T$ = 295 K and (d)-(f) $T$ = 75 K, respectively for the film Si/BPIO.

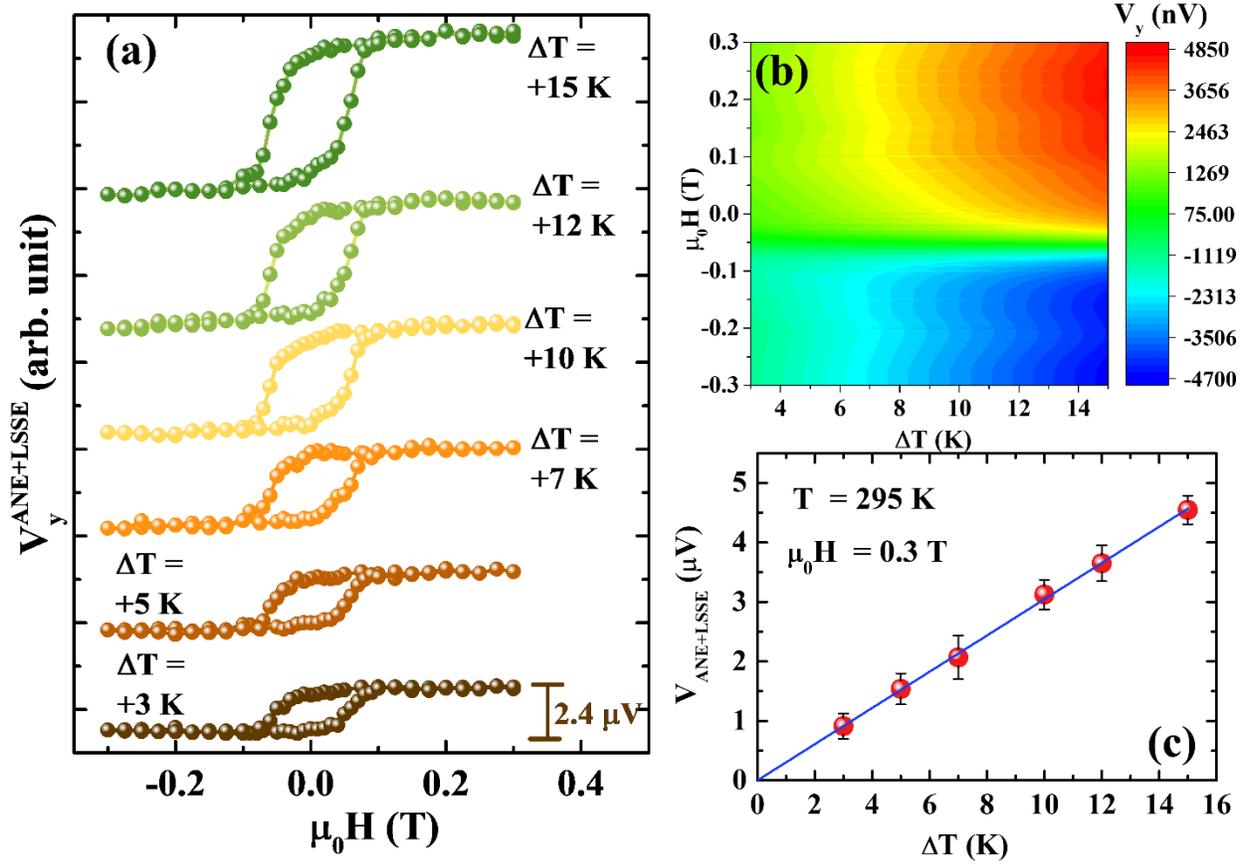

**Figure 5.** (a) Magnetic field dependence of $V_y^{ANE+LSSE}(H)$ across the Pt layer of the Si/BPIO/Pt heterostructure at different temperature gradient intervals between $\Delta T = +5$ and $+15$ K while keeping the sample temperature fixed at $T = 295$ K. (b) Two-dimensional surface plots of $V_y^{ANE+LSSE}(H)$ isotherms for unipolar field scan: $+\mu_0 H_{sat} \rightarrow -\mu_0 H_{sat}$ for the Si/BPIO/Pt heterostructure, (c) the background-corrected voltage $V_y^{ANE+LSSE}(\mu_0 H = 0.3 \text{ T})$ as a function $\Delta T$, which clearly demonstrates a linear behavior.

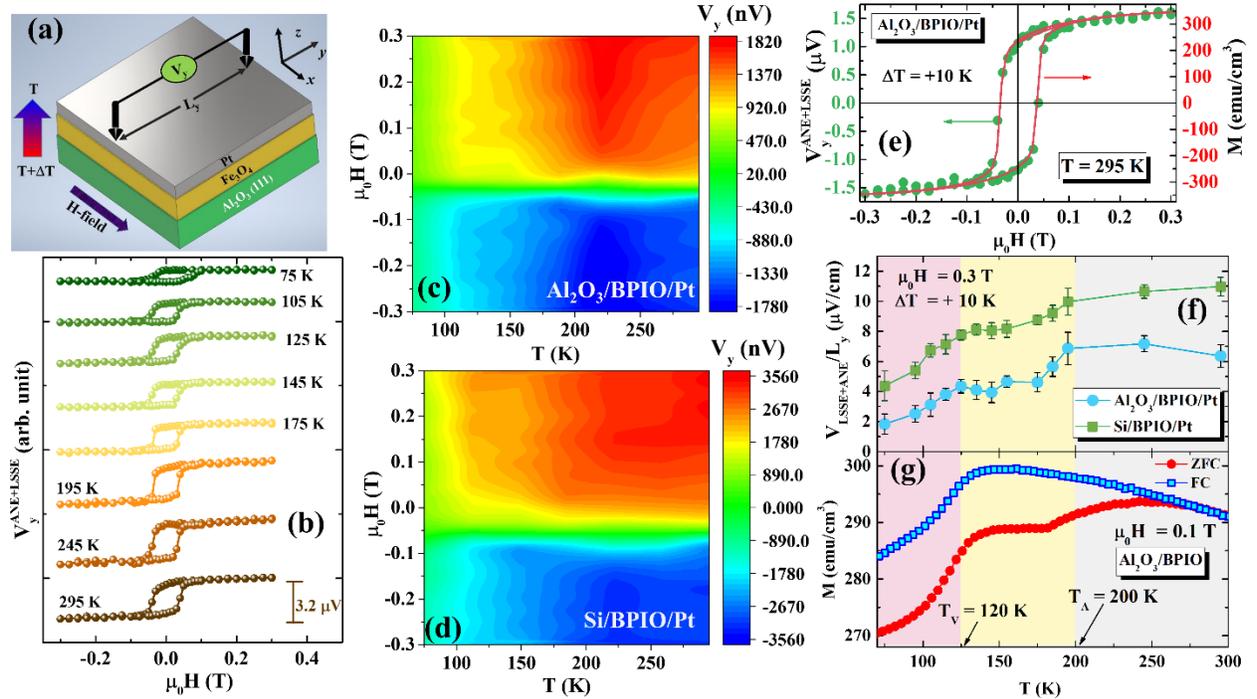

**Figure 6.** (a) Schematic sketch of the LSSE measurement performed the Al$_2$O$_3$/BPIO/Pt heterostructures; (b) magnetic field dependence of the total voltage, $V_y^{ANE+LSSE}(H)$ measured across the Pt layer of the Al$_2$O$_3$/BPIO/Pt heterostructure at selected temperatures between $T = 295$ and 75 K for applied temperature gradient $\Delta T = +10$ K; Comparison of the two-dimensional surface plots of $V_y^{ANE+LSSE}(H)$ for the Al$_2$O$_3$/BPIO/Pt (c) and the Si/BPIO/Pt (d) heterostructures for unipolar field scan: $+\mu_0 H_{sat} \rightarrow -\mu_0 H_{sat}$. (e) Comparison of $V_y^{ANE+LSSE}(H)$ on the left-y scale (symbol) and corresponding $M(H)$ hysteresis loop for Al$_2$O$_3$/BPIO on the right $y$-scale (solid line) at $T = 295$ K. (f) Temperature dependence of the background-corrected voltage normalized with respect to the distance between the contact leads ($L_y$) on the Pt layer, $\frac{V_{ANE+LSSE}}{L_y} = \frac{V_y^{ANE+LSSE}(\mu_0 H = +\ 0.3\ T) - V_y^{ANE+LSSE}(\mu_0 H = -\ 0.3\ T)}{2L_y}$ for the Al$_2$O$_3$/BPIO/Pt and the Si/BPIO/Pt heterostructures; (g) Temperature dependence of the ZFC and FC $M(T)$ for the Al$_2$O$_3$/BPIO bilayer film.

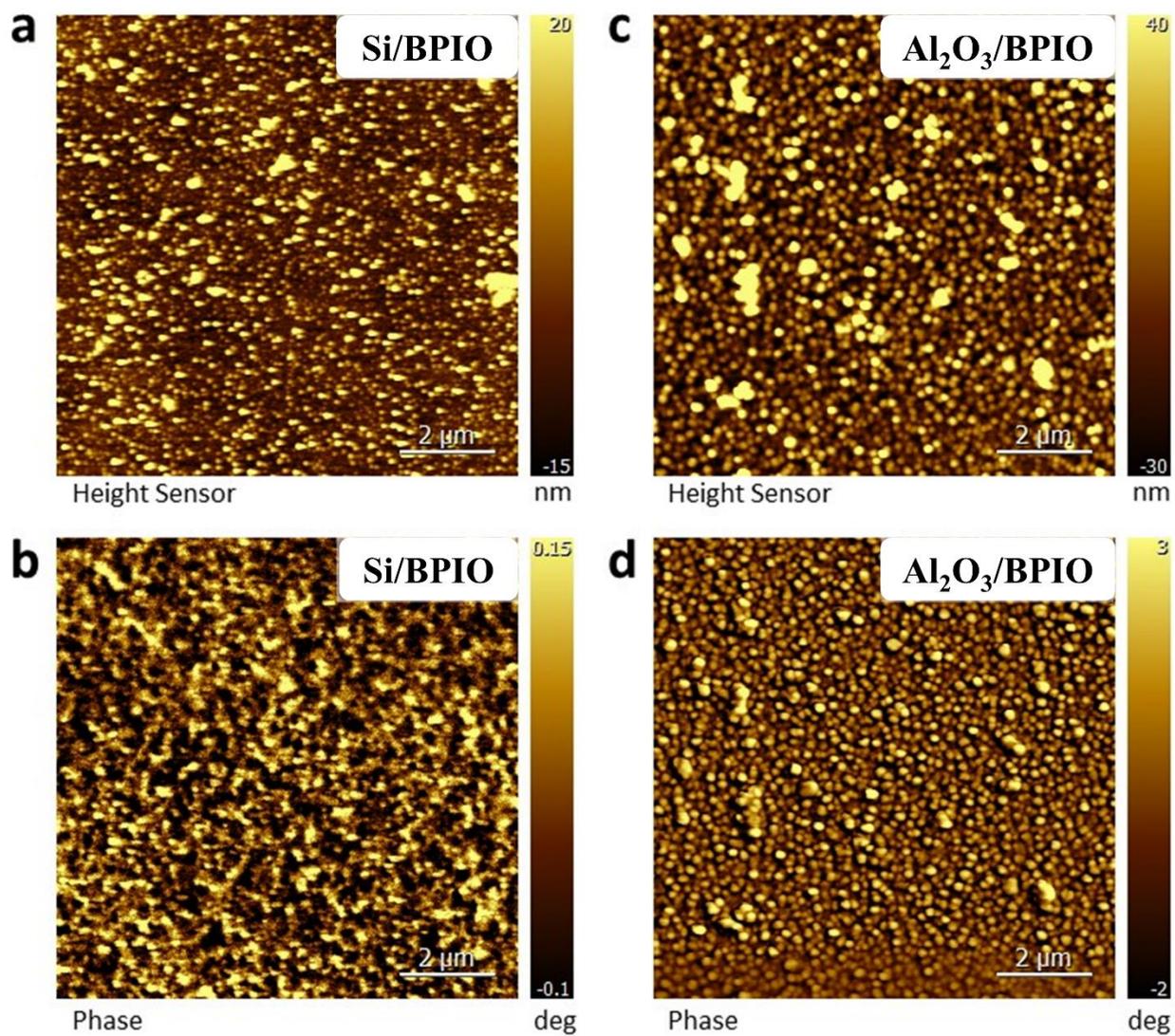

**Figure 7.** (a) and (c) AFM images and, (b) and (d) MFM images of the Si/BPIO and Al$_2$O$_3$/BPIO films, respectively.

# Table of Contents

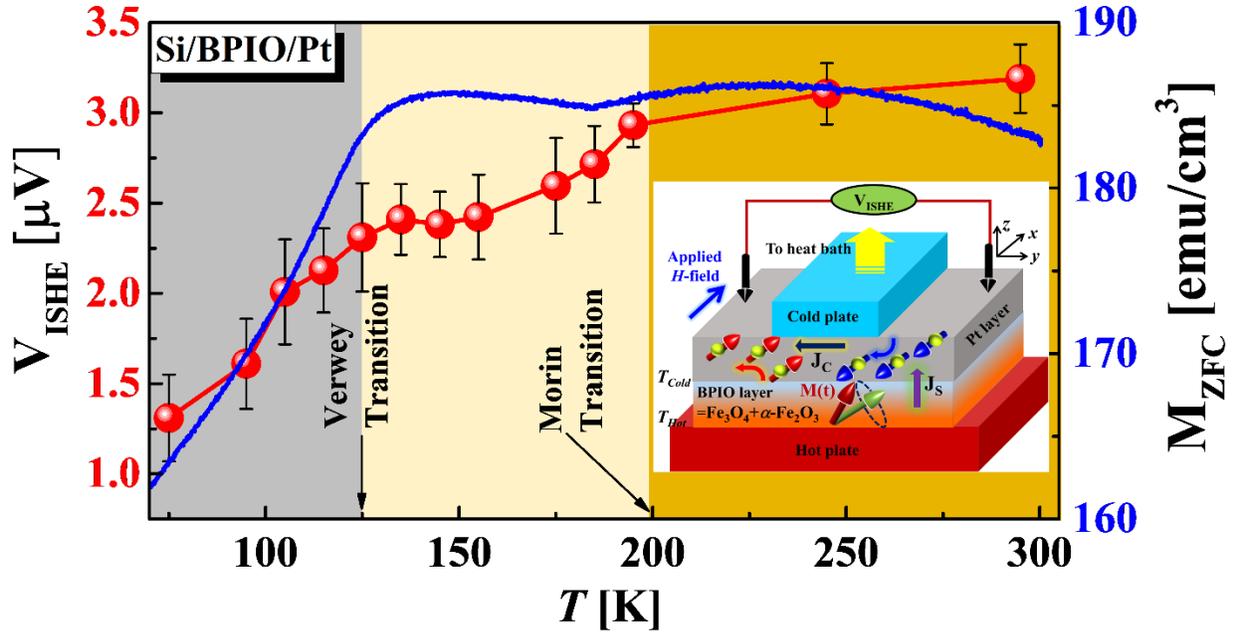

Our new observations of noticeable changes in the longitudinal spin Seebeck voltage around the Verwey transition of the $Fe_3O_4$ phase and the Morin transition of the $\alpha$-$Fe_2O_3$ phase in a biphase iron oxide (BPIO = $\alpha$-$Fe_2O_3$ + $Fe_3O_4$) film demonstrate the important effects of phase transition and coexistence on thermo-spin-transport in BPIO/Pt.